# Advanced Analytics on 3D X-ray Tomography of Irradiated Silicon Carbide Claddings

Fei Xu[1,#], Joshua J. Kane[2], Peng Xu[1,#], Jason L. Schulthess[1], Sean Gonderman[3], Nikolaus L. Cordesa[4]

[1] Idaho National Laboratory, Idaho Falls, ID, United States

[2] Ultra Safe Nuclear Corporation, Seattle, WA, United States

[3] General Atomics, San Diego, CA, United States

[4] Los Alamos National Laboratory, New Mexico, United States

[#]Corresponding authors

Fei Xu: Email: Fei.Xu@inl.gov

Peng Xu, Email: Peng.Xu@inl.gov

**Abstract:**

Silicon Carbide (SiC) ceramic matrix composite (CMC) cladding is currently being pursued as one of the leading candidates for accident-tolerant fuels. To enable an improved understanding of SiC-SiC composite performance, the development of non-destructive evaluation techniques to assess critical defects is needed. Three-dimensional (3D) X-ray imaging, also referred to as X-ray computed tomography (CT), is a non-destructive, data-rich characterization technique that can provide surface and subsurface spatial information. This paper discusses the design and implementation of a fully automatic workflow to detect and analyze SiC-SiC defects using image processing techniques on 3D X-ray images. The workflow consists of four processing blocks, including data preparation, void/crack detection, visualization, and analysis. To implement this workflow the cylindrical axis orthogonal to the volumetric Z-axis is first aligned. A designed automatic algorithm is then used to extract the sub-image containing the material part. Next, the volume from a Cartesian coordinate to a polar coordinate system is transformed, which reduces the image size and, hence, the computational burden by a factor of 60.  Then voids are detected, artifacts are removed, and the defects are labeled. Finally, the connected structures of 3D visualized voids/cracks and statistical results are generated. In this work, three SiC samples (two irradiated and one unirradiated) provided by General Atomics are investigated. The irradiated samples were exposed in a way that was expected to induce cracking, and indeed, the automated workflow developed in this work was able to successfully identify and characterize the crack formation in the irradiated samples while detecting no observed cracking in the unirradiated sample. These results demonstrate the value of automated XCT tools to better understand damage and damage propagation in SiC-SiC structures for nuclear applications.

1. **Introduction**

Accident-tolerant fuels (ATF) are developed for improving the tolerance of the fuel to severe accident conditions such as loss-of-coolant accidents (LOCA) or reactivity-initiated accidents (RIA)[1]. In nuclear reactors, cladding is the thin-walled tube that forms the outer jacket of a nuclear fuel rod. It is critical barrier between the fuel and the reactor coolant and provides structural support to the fuel while retaining fission products. Silicon carbide fiber, silicon carbide matrix (SiC-SiC) composite materials, which consist of a woven layer of nuclear grade SiC fibers overcoated by chemical vapor infiltration (CVI), are investigated as one of the cladding material candidates for AFT because of their outstanding physical and chemical properties[2]. SiC ceramic matrix composites (CMCs) have excellent high-temperature oxidation properties, superior irradiation resistance, inherent low activation, and other superior physical/chemical properties[3]. However, SiC exhibits nonlinear damageable mechanical behavior governed by microcracking within the material under different conditions, such as the stress-state induced during irradiation maintaining hermeticity, and pellet cladding mechanical interaction (PCMI) failure under RIA or other transient conditions in which micro-cracks can develop [4-9].

To design an optimal SiC structure, it is essential to understand the relationship between the damage mechanisms and the microstructure of the SiC CMC material. Micro-cracking is one of the key challenges which impacts cladding performance and guarantees reactor safety. Therefore, quantitative evaluation of crack formation and leak paths are critical to determining SiC CMC in-core performance. Micro-cracks are numerous microstructural features on the material surfaces and subsurface, which spread over the cladding under different scales as shown in Fig. 1 (a)-(c). Currently, destructive analysis like scanning electron microscopy (SEM) and transmission electron microscopy (TEM) is the best method for obtaining micro-crack data from samples. These methods are time-consuming, expensive and compromise the sample preventing further performance evaluations. This highlights the need for improved non-destructive evaluation techniques optimized for micro-crack evaluation and understanding. To address this challenge, the Material Fuel Complex (MFC) of Idaho National Laboratory (INL) has developed post-irradiation examination capabilities to collect non-destructive 3D microstructural data using a ZEISS Xradia 520 Versa X-ray microscope (Fig. 1(d)). This system is used to acquire a wide range of microstructural data over a wide length scale ($10^1 - 10^{-2}$ cm) and can capture both the surface and interior martial structure in 3D[10]. This is accomplished by taking hundreds of X-radiographs on a sample of interest as a function of sample rotation. Then the 3D reconstruction processing will conduct 3D images in a composed stack of material cross-section 2D radiographs. These high-resolution 3D models are then used to explore the sample's architecture and microstructure allowing for better determination of the samples' condition and improved understanding of how the material performance[11-14].

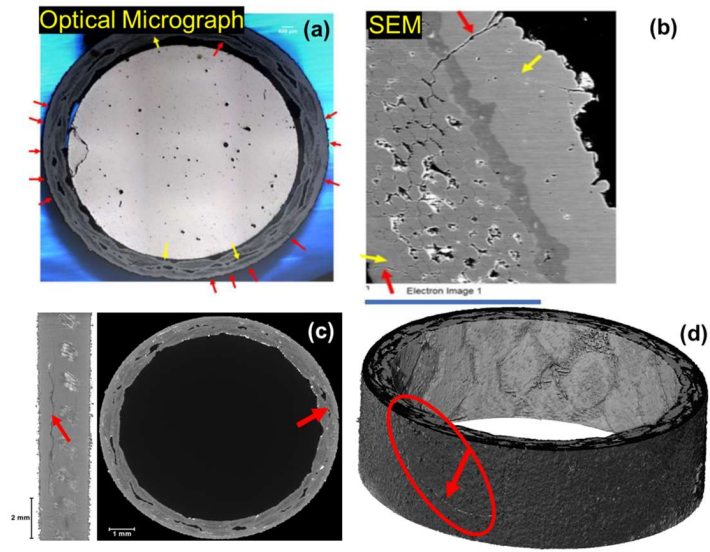

Fig. 1. a) optical micrograph of a fuel cross-section of Sample 1[15] with crack locations identified by the red and yellow arrows; b) high-resolution SEM image of local fuel cladding region of Sample 1[15] with crack locations identified by the red and yellow arrows; c) 2D cross-sectional images in the vertical (left) and horizontal (right) orientations, showing a vertical subsurface crack (red arrows) in the SiC-SiC cladding; d) 3D volume renderings, derived from the X-ray tomograms, of irradiated, defueled SiC-SiC cladding.

In this paper, we developed a fully automatic workflow to detect and analyze cracks using image processing techniques on 3D X-ray images (Fig. 2). Void/crack detection, visualization, and analysis are the three major components, including four processing blocks, noted as $P_1$ to $P_4$. $P_1$ spatially aligned the cylindrical axis orthogonal to the volumetric Z-axis and extracted the smallest possible volume of interest within a Cartesian coordinate system. $P_2$ was applied to unwrap the material annular by transforming the volume from a Cartesian coordinate to a polar coordinate system. $P_3$ detected and labeled the voids/cracks using image segmentation approaches. $P_4$ generated the connected structures of 3D visualized voids/cracks and statistical results. SiC samples provided by General Atomics have been tested in TREAT and evaluated using the proposed workflow.

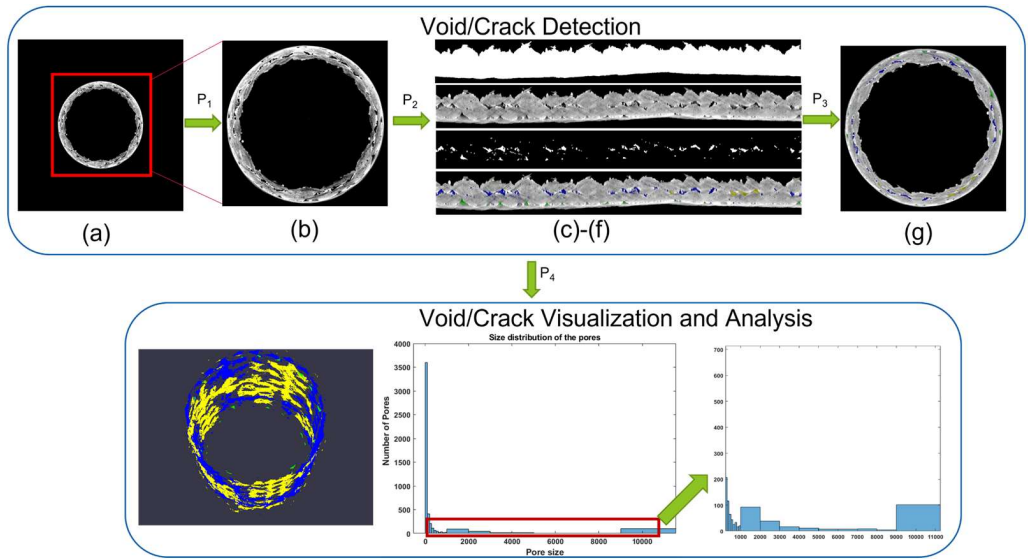

Fig. 2. The proposed workflow. (a) An image from Sample 3 with field of view representing 3024×3064 with resolution 6μm/pixel; (b) sub-image containing material annulus; (c)-(f) represent a spatially transformed coordinate system of Sample 3 showing from top to bottom a binary segmentation of the annulus, original grayscale reconstruction, segmented voids, and color-coded pore classifications. In (f): green denotes the voids/cracks connected directly to the outer surface of the annulus; blue identifies closed voids/cracks that do not touch the inner or outer surface of the annulus; yellow identifies voids/cracks touching the inner surface of the annulus; and red identifies voids/cracks touching the inner and outer surfaces. (g) Rewrapped image frame with corresponding classifications from (f).

## 2. Methods

### 2.1. Experimental data

INL has been the leading national laboratory for research and development on nuclear fuel. This research spans multiple fuel forms and post-irradiation characterization. Modern experiment instruments have been employed to perform transient irradiation testing of advanced fuels, including ATFs at the Transient Reactor Test (TREAT) facility including the irradiated samples that are used in this work. The materials and irradiation parameters of two irradiated fuel rodlets (Sample 1 and Sample 2) with SiC-SiC cladding were shown in Table 1. Sample 1 was noted as SETH-H and Sample 2 was noted as SETH-I in the previous study[15,16]. The melt temperature for SiC is ~2818K(~2445 ℃)[17,18]. The two irradiated samples illustrated the thermal mechanical responses of SiC cladding at high temperatures in transient conditions. Sample 1 received a higher energy deposition when compared to Sample 2. To evaluate pellet-cladding mechanical interaction, Sample 2 was designed with a smaller pellet/cladding gap with the intent of using the thermal expansion of the fuel to provide a rapid strain displacement load on the cladding[15,16]. Pre and post-transient dimensional measurements showed that the rodlet of Sample 1 has a maximum positive diametral strain displacement of ~0.55% near the upper region, while Sample 2 has a strain displacement of ~ 0.65%. Both of the strains are beyond the 0.32% predicted strain in Kamerman et al.[19,20] which would be adequate to initiate cracks in the outer surface of the cladding and allow for crack propagation through the cladding wall. Visual inspection revealed the two irradiated samples maintained intact and rod-like geometries at high temperatures. Moreover, optical microscopy of the sample revealed larger pores in the fuels. Sample 3 is a fresh General Atomics SiC-SiC cladding without fuel.

Non-destructive 3D imaging was performed using the ZEISS Xradia 520 Versa X-ray microscope (Carl Zeiss X-ray Microscopy Inc., Dublin, CA, USA) at INL's Irradiated Materials Characterization Laboratory (IMCL) at Materials and Fuels Complex (MFC). The Nordson DAGE tungsten X-ray source (Aylesbury, UK) was operated with a proprietary high energy filter, 110.9-kVp accelerating voltage, and 111.7-µA target current. All projection radiographs were acquired over 360 degrees of sample rotation, and each radiograph had an average of 20 image frames. The frame size is 3024 by 3064 pixels. The reconstructed 3D slides are shown in Fig.3. Additional detail is listed in Table 2.

Table1 Information of three samples

| Sample | Fuel | Clad | Pellet Clad Gap (µm) | Rodlet Energy Deposited (J/g) | Pulse Width (ms) | Peak Fuel T (C) | Peak Clad T (C) | Objective |
|---|---|---|---|---|---|---|---|---|
| 1 | NE-U3Si2 | SiC | 80 | 528 | 95 | 2050 | 1133-1156 | PCI/Fuel melting |
| 2 | NE-U3Si2 | SiC | 50 | 330 | 100 | 1440 | 831-840 | PCMI |
| 3 | NE-U3Si2 | SiC | | | | | | Fresh sample |

Table 2 3D X-ray data information of three samples

| Sample | Source to the Rotation Axis (RA) (mm) | Detector to the RA (mm) | Pixel Size (μm) | Number of Projections | Cone Angle | # of Slices |
|---|---|---|---|---|---|---|
| 1 | -26.0461 | 245.4407 | 7.1762 | 4,501 | 14.9334 | 1932 |
| 2 | -26.0461 | 245.4392 | 7.1763 | 4,501 | 14.9040 | 1932 |
| 3 | -21.0601 | 234.2374 | 6.1704 | 2,401 | 15.8031 | 1932 |

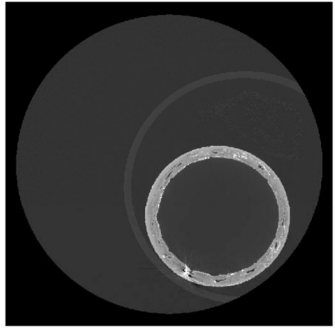
A Slice from Sample 1

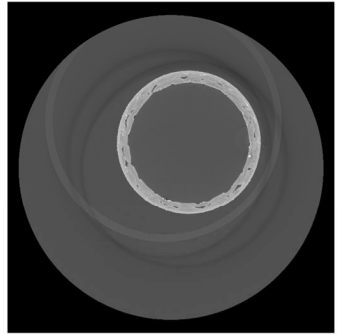
A Slice from Sample 2

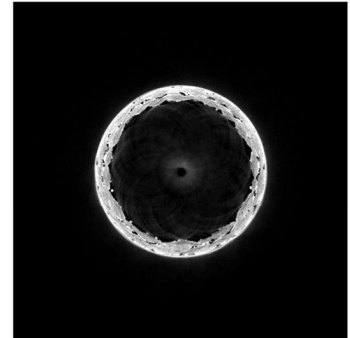
A Slice from Sample 3

Fig. 3. Image slices from the three samples

Although 3D X-ray rendering provided high-resolution and non-destructive data, slices of 3D X-ray exist rotating and drifting during reconstruction which will affect the pores/cracks 3D structure. As shown in Figs. 4 (a) and (c), the cladding circular centers of slice#410 and slice# 1156 from the same sample are (1973,1916) and (2038,1962), separately. Moreover, high-density artifacts were brought in the 3D images (Fig. 4(b)) which impacts the pores detection. Additionally, the material centers vary in different samples

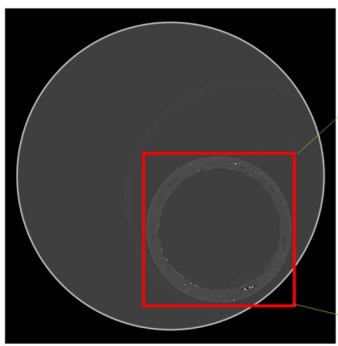
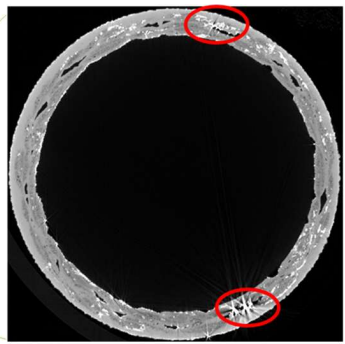
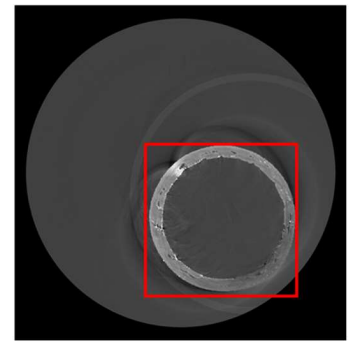

(a) Slice#1156 (3024×3064)   (b) Cladding circular part (1351×1351)   (c) Slice#410

Fig. 4. Existing Challenges

(Fig. 3). To compensate for this a reorientation method was applied to align the cylindrical axis orthogonal to the volumetric Z-axis and an automatic algorithm was designed to extract the sub-image containing the material part.

## 2.2. Reorientation and Region of Interest (ROI) extraction

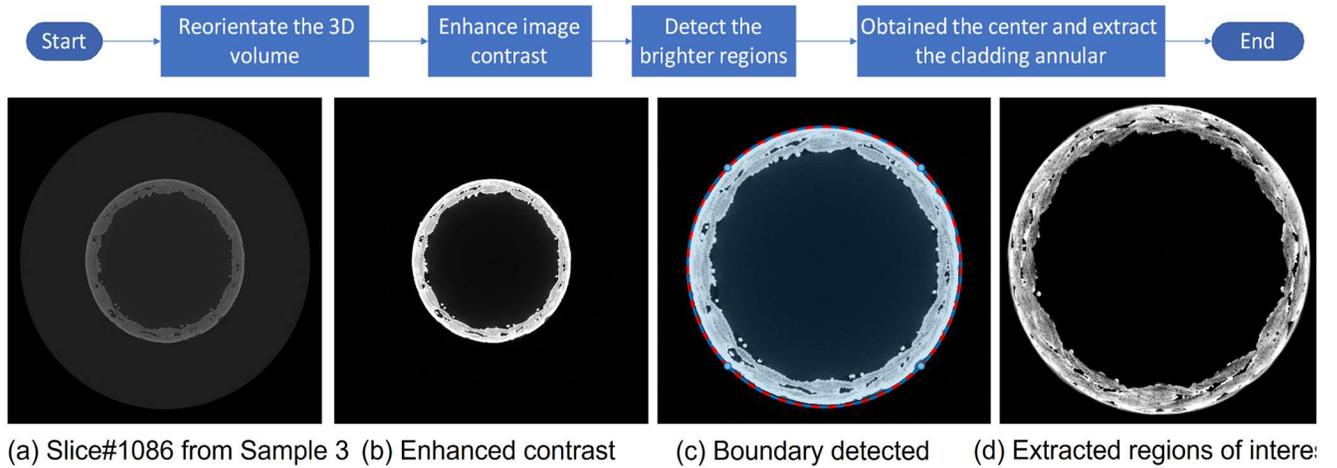

(a) Slice#1086 from Sample 3  (b) Enhanced contrast   (c) Boundary detected   (d) Extracted regions of interest

Fig. 5. Reorientation and ROI extraction. (a) original image with size 3024×3064; (b) contrast enhancement; (c) zoomed-in cladding boundary detection result; (d) exacting the cladding circulars after reorientation with image size 1351×1351

As shown in Fig. 2, the process of $P_1$ was to align the volume and extract the cladding circulars. The flowchart of the process and the corresponding intermediate results are shown in Fig. 5. In the reorientation process, we applied an algorithm to align the cylindrical axis to the volumetric Z-axis. The algorithm has two inputs, the 3D volume and a 2D matrix consisting of $(x_i, y_i, i)$, namely *Center*, where $(x_i, y_i)$ is the cladding center of the $i$th slice. The cladding center was generated automatically through the circular boundary detection results (Fig. 5(c)). The value of each pixel is named intensity in the image, and the brighter regions indicate higher intensity. As shown in Fig. 5(b), the cladding appears much brighter than the other regions. We detected the material circular by using a global threshold segmentation method[21]. The threshold method sets the pixel value to be 1s if the image intensities are greater than an adaptive threshold $T_0$; otherwise, assigns 0s. The center of the cladding circular was calculated by using the detected outer boundary of the cladding. In the study, we created a list of centers for the first and last slices, named *Center*.

The reorientation process contains the three main steps: 1) Determine misorientation angles along *x*-axis and *y*-axis from *Center* by generating the slopes on *x* and *y*-axis along the *z*-axis (slice number); 2) assuming the orientation and drift changing are symmetric, generate the transform matrix linear translation on the first half volume steps by [*mean(x)*, *mean(y)*, *the number of slices/2*] and the second half volume by –[*mean*(x), *mean(y)*, *the number of slices/2*] along *x*-axis, *y*-axis and *z*-axis respectively. *mean(x)* and *mean(y)* are the mean x-axis and y-axis values in *Center*; 3) apply a spatial transformation with the translation matrix to compute the corresponding locations in the original 3D volume transform subscript space for each location in the reorientated space as shown in Figs. 5 (c) and (d).

After reorientation, the cladding part for each slice is automatically detected and the volume only containing the cladding circular in each slice is cropped. The original volume size of Sample 3 is 17GB, but the sub-volume with the cladding is 2GB, which reduced the storage more than 8 times.

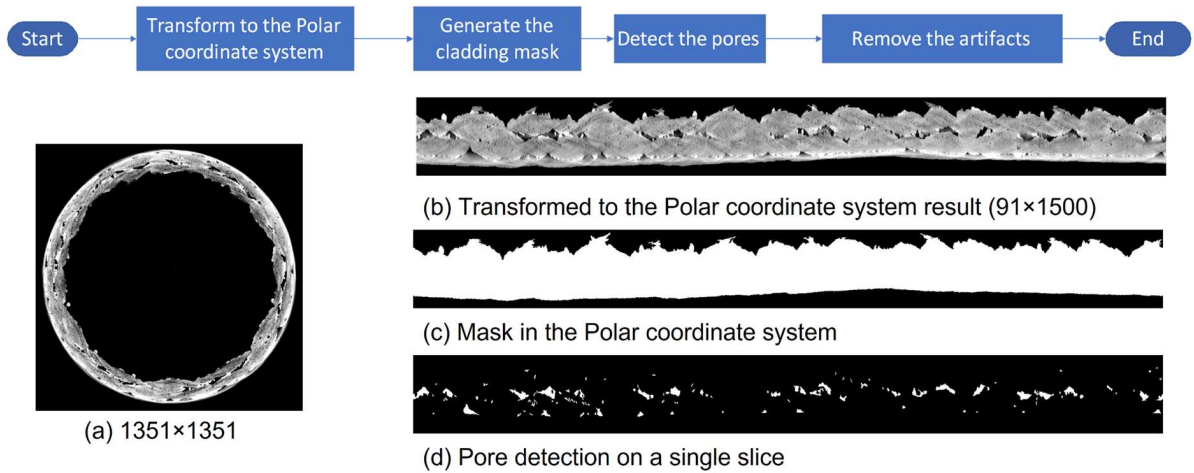

Fig. 6. Voids detection.

## 2.3. Voids detection

Fig. 6 showed the process of voids detection. First, the Cartesian coordinate system is transformed into the Polar system to reduce the computational complexity. Since the cladding material is an annular shape, the corresponding elements of the two-dimensional Cartesian coordinate arrays *x* and *y* are transformed into polar coordinates *theta* and *rho*. The radius of the inner and outer boundaries of the cladding are nearly 550 and 650 pixels from the center. The thickness of the cladding is less than 100 pixels. 1500 angles with the radius range covering the inner and outer boundaries of the cladding in the Polar system are generated. The transformed image size is 91 × 1500 as shown in Fig. 6(b), which reduced the computational size by nearly 7 times.

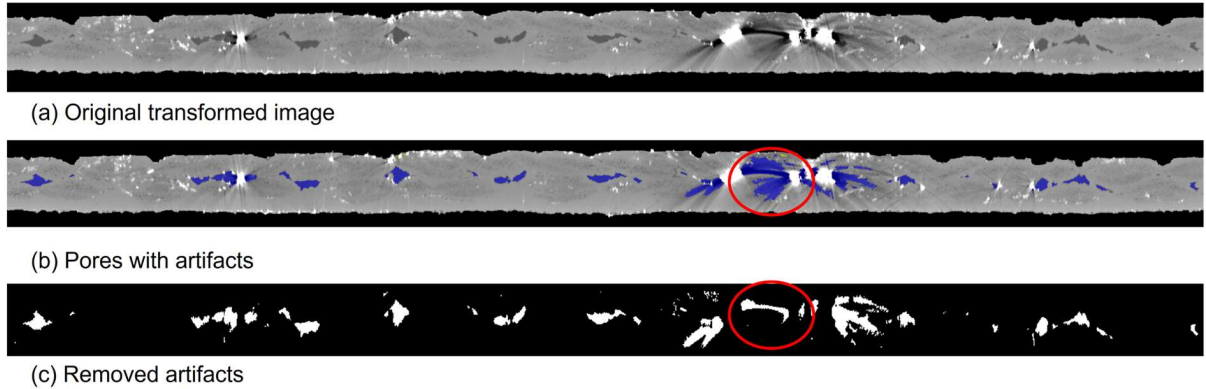

(a) Original transformed image

(b) Pores with artifacts

(c) Removed artifacts

Fig. 7. Remove the shader Voids detection.

After transformation, the voids in the cladding regions are detected (Fig. 6(c)). The voids appeared darker than the other regions in the cladding. Noises is removed and the voids are detected using the adaptive threshold segmentation method[21] after adjusting the contrast of the image. However, the shadows, as marked in the red circles in Figs. 7(a) and (b), resulted from the high-density metal during X-ray scanning[22,23]. The shadows are inhomogeneous with and surround the voids and are not easily distinguished from the intensities, which will result in misdetection by the threshold segmentation method. , are . To solve the problem, the intensities' standard derivation for each connected object are calculated; and if the standard derivation is greater than $d_1$, an adaptive local threshold value in the connected void is generated, and the pixels with intensities greater than the value are removed (Fig. 7(c)).

2.4. Visualization and analysis

As studied previously by Schulthess, *et al*[15], the rapid thermal expansion of the fuel pellet contacting the cladding and the heat transfer to the cladding causing a thermal stress gradient across the cladding wall resulted in cracks in Sample 1 and Sample 2. To classify the pore and crack structures, the pores/cracks into four categories were created: the pores/cracks inside of cladding, the pores/cracks adjacent to the outer boundary, the pores/cracks adjacent to the inner boundary, and the pores/cracks connecting the inner and outer boundary, noted as $c_1$-$c_4$, respectively. The 3D structural information of the voids and cracks ($c_4$) was generated by the built-in function regionprops3 in Matlab[24,25].

**Porosity statistical analysis**. The overall number of voids and cracks in the three samples is shown in Table 3. Sample 1 had the greatest number of voids, but Sample 2 had the most porosity ratio. Sample 1 and Sample 2 contained nearly or over twice the number of voids than the fresh Sample 3. As shown in column $c_4$ in Table 4, no crack that crossed the inner and exterior boundaries of cladding was detected in Sample 3, while Sample 1 was founded the greatest number of individual cracks, and Sample 2 had the largest cracks. From the aspect of voids' locations, nearly all of the voids (>99%) of Sample 3 were inside the cladding ($c_1$). While most voids of Sample 1 and Sample 2 were cracks that crossed the inner and exterior boundaries of the cladding ($c_4$) and microcracks ($c_2$), which matched the characterization observation in the previous study[15]. The crack structures of Sample 1 and Sample 2 are shown in Figures 8 and 9. The darker regions in the figures are voids. The colored components are the cracks. The properties of the cracks were generated as shown in Table 4, including the volume size of the crack, and principle axis length. The principle axis length is the length of the major axes of the ellipsoid that have the same normalized second central moments as the region, as a 1-by-3 vector. The data in Table 4 revealed that all the cracks have longer lengths in the vertical direction, especially crack 1 of Sample 2. Sample 2 had a less number of cracks, but each crack was larger than the cracks in Sample 1.

Table 3 Overall statistics of Voids and Cracks

| Sample # | Volume size | Cladding size (mm$^3$) | # of voids | Porosity ratio | $c_1$ | $c_2$ | $c_3$ | $c_4$ | # of void with size >0.001mm$^3$ | # of void with size >0.01mm$^3$ | # of separated cracks |
|---|---|---|---|---|---|---|---|---|---|---|---|
| 1 | 91×1500×1111 | 55.39 | 97446 | 5.73% | 26% | 40% | 1% | 33% | 124 | 97 | 8 |
| 2 | 91×1500×1865 | 92.98 | 84333 | 6.96% | 23% | 11% | 2% | 63% | 84 | 66 | 2 |
| 3 | 220×1500×1931 | 85.23 | 4828 | 3.01% | 99% | 1% | 0% | 0 | 135 | 55 | 0 |

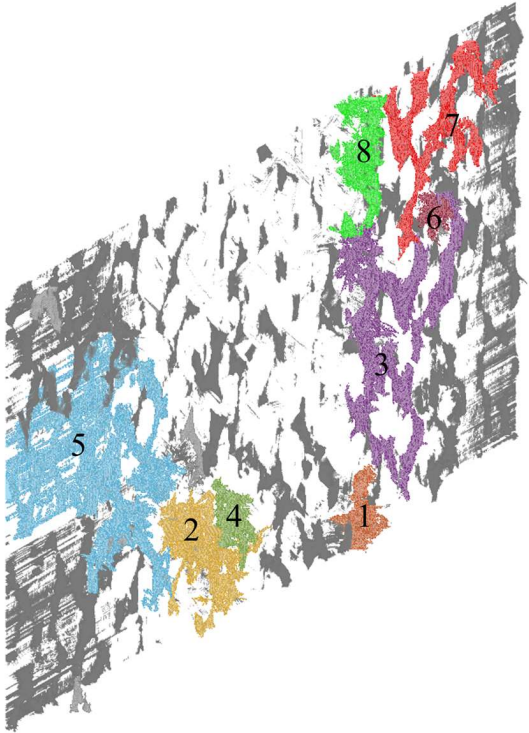

Fig. 8. Crack structure in unwrapped Sample 1

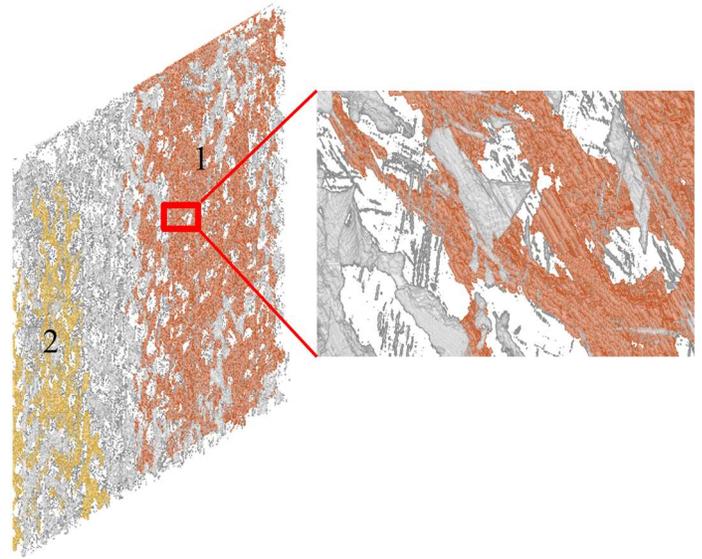

Fig. 9. Crack structure in unwrapped Sample 2

Table 4 Overall statistics of Voids and Cracks

| Sample# | # of Crack | Volume size (mm$^3$) | Principle axis length (mm) | | |
|---|---|---|---|---|---|
| Sample1 | 1 | 0.04 | 1.41 | 0.81 | 0.43 |
| | 2 | 0.12 | 3.19 | 1.28 | 0.54 |
| | 3 | 0.28 | 5.50 | 2.71 | 0.37 |
| | 4 | 0.04 | 1.76 | 0.60 | 0.48 |
| | 5 | 0.32 | 5.11 | 3.21 | 0.56 |
| | 6 | 0.01 | 0.69 | 0.55 | 0.43 |
| | 7 | 0.16 | 3.25 | 2.79 | 0.26 |
| | 8 | 0.08 | 2.88 | 0.86 | 0.40 |
| Sample 2 | 1 | 3.23 | 17.38 | 6.97 | 0.72 |

|   | 2 | 0.87 | 13.51 | 3.66 | 0.63 |

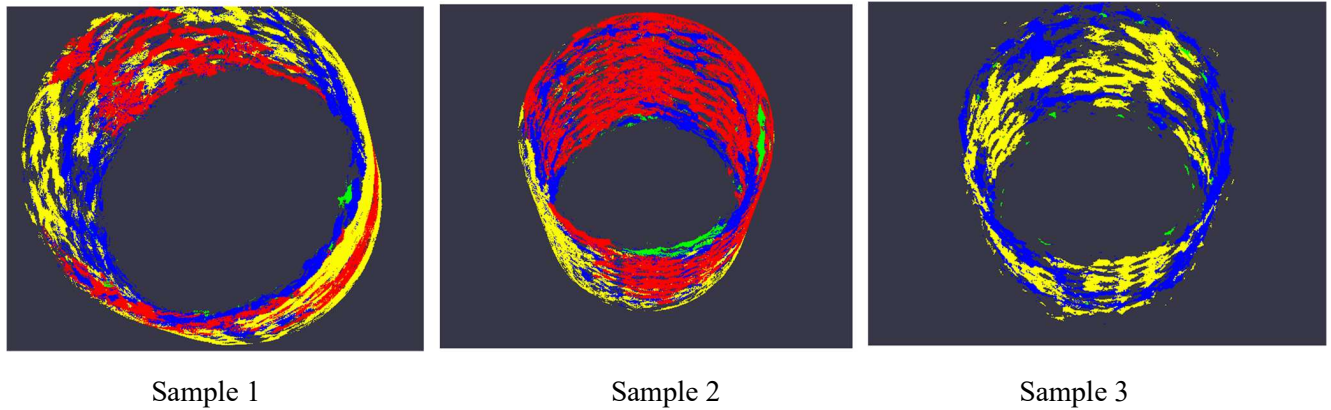

| Sample 1 | Sample 2 | Sample 3 |

Fig. 10. Crack structure in unwrapped Sample 1, Sample 2 and Sample 3.

**PIE data analysis**. The SEM characterization of Sample 1 indicated that in the transverse section surface observed 14 radial cracks and 3 circumferential cracks, while 2 radial cracks and 3 axial cracks were found in the longitudinal section. The cracks was found interconnected through the cladding thickness from the cross-section.

Optical microscopy of Sample 2 revealed more radial and circumferentially cracks in the SiC cladding than in Sample 1[15]. On the sample surface, 31 radial cracks and 10 circumferential cracks in the transverse section, and 15 radial and 0 axial cracks in the longitudinal section are observed. At least 6 of the cracks were found to propagate through the wall thickness of the cladding. Many large pores ranging in size from ~300 to ~400 μm were found. Interconnected cracks appeared along with branching cracks emanating from the main crack. Moreover, a crack network was observed on the inner surface, which may be from fabrication. More SEM investigation revealed more branching cracks and widespread micro cracks. All the findings from the sample surface matched the X-ray 3D results.

This is the first work to illustrate the 3D X-ray nondestructive capability of defects and cracks in SiC CMC cladding after transient safety testing at INL. The degree of cracking is proportional to the transient strain the cladding experienced in the TREAT experiments. The statistics of the defects and characteristics can inform the mechanical responses of SiC cladding in accidents, and can be later correlated to manufacturing for processing and performance improvement.

3. Conclusion

This work proposed a workflow to automatically process 3D X-ray data, which integrated the preprocessing, object detection and visualization, and demonstrated the applicability of X-ray tomographic imaging capability in rapidly supporting post-irradiation examinations of non-fuel components.

Three advantages of the proposed workflow are listed as: a) the automatic ROI extraction and transformation from the Cartesian coordinate system into the Polar system processes reduced computational size by nearly 60 times; b) simple threshold-based segmentation handled the automatic pores detection efficiently;  c) 3D visualization voids and cracks analysis revealed the overall porosity condition, location, and structure of the voids and cracks which provided critical quantitative information to evaluate the material condition and performance and eventually assist the decision-maker to determine the material lifetime.

The experimental results demonstrated that Sample 1 (lower strain) and Sample 2 (higher strain) have cracks establishing a continuous path between the interior and exterior surface of the annulus breaking the desired hermiticity of the material, and no such cracks were observed in Sample 3 (fresh material). These results demonstrate the applicability of X-ray tomographic imaging capability in rapidly supporting post-irradiation examinations of non-fuel components X-ray tomography with advanced visualization methods and automated defect analysis. The workflow can be applied to other similar 3D X-ray data analysis tasks.

Moreover, the current X-ray instrument capability at INL can provide the image quality with length scales 10 to $10^{-2}$ cm which is not enough to observe the microcracks whose width or length is under that. Due to the resolution limitation, the crack or pore whose size is less than 35 µm2 (5 pixels) cannot be detected in this study.  In the future, we will apply the super-resolution models to generate high-resolution images to analyze microcracks on the volume, and investigate more advanced analysis on SEM, STEM, and EELS data to reveal the relationship between chemical composition and crack formation.

## 4. Acknowledgements

This research is being performed using funding received from Advanced Fuel Campaign. The studied samples are provided by General Atomics.

## 5. Disclaimer

Authors have no known competing financial interests.

This information was prepared as an account of work sponsored by an agency of the U.S. Government. Neither the U.S. Government nor any agency thereof, nor any of their employees, makes any warranty, express or implied, or assumes any legal liability or responsibility for the accuracy, completeness, or usefulness of any information, apparatus, product, or process disclosed, or represents that its use would not infringe privately owned rights. References herein to any specific commercial product, process, or service by trade name, trademark, manufacturer, or otherwise, does not necessarily constitute or imply its endorsement, recommendation, or favoring by the U.S. Government or any agency thereof. The views and opinions of authors expressed herein do not necessarily state or reflect those of the U.S. Government or any agency thereof.

## 6. Data availability statement

The data that support the findings of this study are available from the corresponding author upon reasonable request.